# WHO CREATES THE TIME: NATURE OR HUMAN?

**Sergey B. Kulikov***

Tomsk State Pedagogical University
Tomsk, Russian Federation

## ABSTRACT

The aim of this article is to defend the thesis that analysis of time meaning within history and philosophy of natural sciences and philosophical anthropology allows making clear the basis of human being. It's opened the opportunity of constructing special model of general understanding of time as a creation of nature or as a creation of human. Two main methods are used: comparative analysis and hermeneutics.

Article presents the discussion of following results. Orientation on discretization and virtual nature of cultural interaction, or orientation on mutual tension of limits of cultural and historical process allows connecting philosophy of natural sciences and philosophical anthropology with system of physical categories: energy, weight, distance, etc. It finds an application as in the physical and mathematical sphere so in the field of humanistic studies. The general conclusion made is that neither nature nor human solely creates the time. Time is an imaginary phenomenon connecting human activity and natural processes in the limits of human consciousness.

## CLASSIFICATION

APA:   2340, 2630, 4010

*Corresponding author, $\eta$: kulikovsb@tspu.edu.ru; +7 382 2 522 382;
 Faculty of University-wide Disciplines, Kievskaja St. 60, Tomsk, Russia

# INTRODUCTION

## BACKGROUND

An article is clearing up the ratio of meanings of time in a context of philosophy of physical and mathematical natural science and philosophical anthropology. Relevance of the matter is caused at least from two parties. Firstly, cultural development of scientific knowledge assumes philosophical justification [1; pp.259-260]. According to Kant, time concept is the cornerstone of mathematical knowledge and because of that it allows developing conditions of possibility of understanding of physical processes [2; pp.84-88]. Therefore the analysis of understanding of time opens the specific direction of philosophical justification of physical and mathematical natural science. Secondly, the time concept sends not only to area of natural processes but also their sense concerning a being perspective in general and questions of human being in particular [3; p.27, 3; p.451]. Therefore the analysis of time meaning allows clearing up the basis of human being and constructing special model of its philosophical and anthropological understanding.

## APPROACHES AND METHODS

The analysis of question of time meaning as relation "Nature vs. Human" allows to reveal specifics of philosophy, natural sciences and anthropology in structure of ontological knowledge as itself. It is revealed ways of creation of the general theory of nature development and also human interaction. Thereby, we can answer the question is whether nature or human creates the time.

During research were applied methods of the comparative analysis and a hermeneutics. These methods allow comparing the features of interpretation of time within fundamentals of physical and mathematical natural sciences and philosophical anthropology. So, it could be constructed the general model of interpretation.

# CLASSICAL POSITION: NATURE "CREATES" THE TIME

There are two most important positions in the history of time ontology: classical natural philosophical approach (Aristotle, Hegel, etc.) and accenting on human being approach (Husserl, Heidegger, etc.). Bulks of modern theories, both in physical and mathematical natural sciences, and within humanitarian thought are in fact continuation of these directions. For example unlike traditional views in the theory of strings total more than three dimensions of space-time continuum. However, in its basis the theory of strings is a kind of natural philosophy. In a counterbalance to it modern historical issues (for example by H. White) are in many respects obliged to philosophical and anthropological ideas of the first half of the XXth century.

Aristotle analyzes time in categories of existence and nonexistence as natural (physical) phenomena [4; pp.145-156]. As starting position fixing of essentially problematical nature of disclosure of the nature of time acts [4; p.145]. The analysis of traditional for its era concepts leads Aristotle to a number of the important conclusions. Firstly, time is not movement. It cannot "move" (in itself) regarding some necessary way of measurement of mobility for it, but such way coincides with time and it conducts to need of time interpretation through concepts of immovability and invariance [4; p.147]. Secondly, time also has to possess variability and mobility signs [4; p.147]. Therefore time possesses both immovability signs and lines of mobility. Time in itself cannot be divided but in the same relation has to be subdivided into separate parts. The nature of time is inconsistent and demands the special understanding allowing eliminating the revealed contradiction. In this regard Aristotle believes that time is a measure of the movement expressed in the numerical relation to any possible movement and change [4; pp.149-150]. It is issued the understanding of time as the paradoxical phenomenon acting both as complete image and internally divided phenomenon. The paradox of time demands representation it as the formation similar to points of mathematical lines. It is expressed that division of time is applied in cases of ensuring understanding of the message on some event [4; p.155]. Thus, time is inconsistent (uniform and multiple) forming is in itself motionless and invariable and acts as numerical expression of occurring movements and changes, and time also appears in the form of a complex of pragmatic rules of ordinary language practice. All this testifies Aristotle understands time in a relative separation from real processes in the material nature, pulls together it with area of human relationship and gives the grounds to identify it with epiphenomenon of public life.

Special interpretation of time within history and philosophy of natural sciences is represented by Hegel. Hegel correlates time to space including it in process of dialectic formation of Absolute Spirit at an embodiment stage as initially Absolute Idea within an alter-being of this Idea ordering of one of the parties of a material world [5; p.51]. Thus, the general understanding of time coincides with two moments: 1) contradiction to space; 2) the discretization forming in total of points of gap independent images namely the line and a surface (i.e. the plane).

Hegel reproduces Aristotle's idea of rather paradoxical character of the nature of time, but Hegelian position only in a form coincides with the point of view of Aristotle [5; p.52]. Time reveals in categories of existence and nonexistence as internally inconsistent phenomenon, by permission of this contradiction the direction opposite to the Aristotelian line of thought acts. Hegel believes that time in itself coincides with physical process (duration) and puts forward the thesis about limited character of its paradoxicality, abnegation of this paradoxicality at higher steps of development of the nature [5; pp.54-55]. Time is a pure form of the organization of representation of natural processes. In this regard processes are constituted by the moments of distinction which reveals as infinite division and uniform reproduction of reference points of change of an ontological situation. Being is represented within the concepts sending to possibility of some number of positions and states (past, present or future) and reveals as correlated to a necessary scale of transitions between these positions and states. The scale possesses inconsistent properties and cannot be understood with accuracy as cash actually or absent in it. It is connected with Hegelian interpretation of time in itself as "being disappearances in nothing and nothing in being" [5; p.56].

On the basis of the analysis of classical approaches to representation of time it is possible to draw some general conclusions. Firstly, the nature of time has internally inconsistent (paradoxical) character and consists in unity of its existence and nonexistence. Secondly, time is identified with imaginary movement in the Greek classics and also ideal (formal) process of formation of the nature in the German classical thought. Thirdly, the general understanding of time is connected with its "mathematic" character, i.e. formality and universalistic meaning as regulation of language practice, according to Aristotle, and also the organization of quantitative parameters of intelligence of natural processes, according to Hegel.

These conclusions allow correlating natural philosophical interpretation of time to its philosophical and anthropological understanding. The foundation is laid by Heidegger who developed Husserl's ideas in a context to consecutive criticism of bases of science in Modern Times. Husserl in his criticism came to idea of replacement in physical and mathematical natural sciences of a direct reality of the sensual world a complex of geometrical methods of idealization of the phenomena of the material nature [6; pp.8-9]. Heidegger developed that researches and opened special understanding of time.

## NON-CLASSICAL POSITION: HUMAN EXISTENCE AS THE "CREATOR" OF TIME

### PHILOSOPHICAL POINT OF VIEW

According to Heidegger, basis of conceptualization of time is the existential analytics of presence, i.e. being of human in the world and also conditionality of scientific knowledge (about the human, society and nature) as basic intentions of its existence [3; p.29]. According to Heidegger, the nature of time coincides with two versions of interpretation: "ordinary" which sense corresponds to irreflexively accepted assumption of identity of time and direct fixing of course of natural processes [3; p.475] and also the "ecstatic and horizon" interpretation meaning a reflection of the intention of an ordinary understanding of time in the form of irreversible alternation of the moments of change of states within surrounding reality. The second interpretation assumes that daily intention sets a way of understanding of the world and Heidegger considers that time nature in principle corresponds to this ability of the human to be present, be in the world, in a literal word meaning proceeding from itself [3; p.475].

Thus, the classical understanding of time and modern interpretation correspond, firstly, as intention on correlation allocation between natural processes and the sphere of their knowledge, and, secondly, as the form of intension on time restriction with area of human being. Aristotle and Hegel reveal a paradoxicality, the general formality and universality of time regulating (to all appearances and through a prism of language norms, according to Aristotle, and also it is ideal, according to Hegel) natural processes. Heidegger opens irreflexivity of this concept at all and believes that time belongs only to the sphere of being of the human in the world, but not to world being as itself.

It is easy to see that the philosophical and anthropological concept complicates exact compliance of time of area of natural sciences. In a necessary way time cannot be included in all completeness of own sense in a framework of the valid process of natural-science research. The reality of such process has to be called into question in borders of actual understanding of time. But we can show the way to solve this problem.

### INTERDISCIPLINARY POINT OF VIEW: PHILOSOPHY OF TIME WITHIN PHYSICS?

We take the principled stand that there are some fundamental prerequisites in following relations:

(1) idea of discretization of cultural and historical process;

(2) virtual nature of interaction of consciousness limits in philosophical and scientific knowledge (elements of limitation are found in real history and are in the oppositional relation to last and (or) hypothetically possible future moments) [7; pp.11-44].

Our position sends to a complex of philosophical provisions but also can be formulated in natural-science categories of a time point as $t$, energy as $E$, $m$ as body mass, and also distances as $r$ (enter the classical theory of gravity (Newton) and relativistic mechanics (Einstein)). There is possible a special interpretation of time. Time connects action of the phenomenon or its *energy* with some kind of *weight*, the historical importance of a separate limit of consciousness through distance analogue as intensity of cultural and historical space, i.e. oppositional nature of interaction of intentions.

So, the complex of philosophical and physics ideas which allow finding a way of rational justification of natural sciences comes to light coordinating philosophical and anthropological understanding of a ratio of time and being, and also elements of physical science.

## CONCLUSION

Evolution of representations results of time meaning within history and philosophy of natural sciences (esp. within history and philosophy of physics) leads to the problematical correlation of time to natural processes. The meaning of time assumes its paradoxical, ideal and universal character; reality and time correspond only to the world of phenomena of human existence.

We have found one of the ways of solving this problem. Orientation on the rules of discretization and virtual nature of interaction, or orientation on mutual tension of limits of cultural and historical process allow connecting philosophy with system of physical categories (energy, weight and others), which find application as in the physical and mathematical sphere so in the field of humanitarian researches.

Thus one can conclude that neither nature nor human create the time. Time is an imaginary phenomenon connecting human activity and natural processes in the limits of human consciousness.

## TKO STVARA VRIJEME: PRIRODA ILI ČOVJEK?

S. B. Kulikov

Državno učiteljsko sveučilište Tomsk
Tomsk, Rusija

### SAŽETAK

Rad obrazlaže tezu kako analiza značenja vremena u povijesti i filozofiji prirodnih znanosti te u filozofskoj antropologiji razjašnjava temelj ljudskog življenja. To otvara mogućnost konstruiranja posebnih modela općeg razumijevanja vremena kao tvorevine prirode ili kao ljudske tvorevine. U radu se koriste dvije metode: komparativna analiza i hermeneutika.

U radu je razmotreno sljedeće: orijentiranje na diskretizaciju i virtualnu prirodu kulturalnih međudjelovanja, orijentiranje na uzajamno rastezanje granica kulturalnih i povijesnih procesa koji su omogućili povezivanje filozofije prirodnih znanosti i filozofske antropologije sa sustavom fizičkih veličina poput energije, mase, udaljenosti itd. Nađene su primjene u područjima fizike, matematike i humanističkom području. Opći je zaključak kako ni priroda niti čovjek zasebno ne stvaraju vrijeme. Vrijeme je imaginarna pojava koja povezuje ljudsku aktivnost i prirodne procese na granici ljudske svijesti.

## KLJUČNE RIJEČI

ontologija vremena, značenje vremena u prirodoslovlju i matematici, značenje vremena u filozofskoj antropologiji, stvoritelj vremena